\documentclass[12pt,preprint,showpacs,preprintnumbers]{revtex4}
\usepackage{amssymb}
\usepackage{amsmath}
\usepackage{graphicx}
\usepackage{dcolumn}
\usepackage{bm}
\usepackage{epsfig}
\usepackage[T1]{fontenc}
\usepackage{ae,aecompl}
\usepackage{color}

\begin{document}
\title{Three-dimensional discrete Boltzmann models for compressible flows in and out of equilibrium}
\author{Yanbiao Gan$^{1,5}$, Aiguo Xu$^{2,3}$\footnote{Corresponding author. Email address: Xu\_Aiguo@iapcm.ac.cn}, Guangcai Zhang$
^2$, Huilin Lai$^{4,5}$}
\affiliation{$^1$North China Institute of Aerospace Engineering, Langfang 065000,
 P.R.China \\
$^2$National Key Laboratory of Computational Physics, \\
Institute of Applied Physics and Computational Mathematics, P. O.
Box 8009-26, Beijing 100088, P.R.China \\
$^3$Center for Applied Physics and Technology, MOE Key Center for High Energy Density Physics Simulations, College of Engineering, Peking University, Beijing 100871, P.R.China \\
$^4$School of Mathematics and Computer Science,
Fujian Normal University, Fuzhou 350007, China\\
$^5$Fujian Key Laborotary of Mathematical Analysis and Applications,
Fujian Normal University, Fuzhou 350007, China}
\date{\today }

\begin{abstract}
We present a series of three-dimensional discrete Boltzmann (DB) models
for compressible flows in and out of equilibrium. The key formulating technique is the
construction of discrete equilibrium distribution function through
inversely solving the kinetic moment relations that it satisfies.
The crucial physical requirement is that all the used kinetic moment
relations must be consistent with the non-equilibrium statistical mechanics.
The necessity of such a kinetic model is that, with increasing the complexity of flows,
the dynamical characterization of non-equilibrium state and the understanding of the constitutive
relations need higher order kinetic moments and their evolution.
The DB models at the Euler and Navier-Stokes levels proposed by this scheme are validated by several
well-known benchmarks, ranging from one-dimension to three-dimension.
Particularly, when the local Mach number, temperature ratio, and pressure
ratio are as large as $10^2$, $10^4$, and $10^5$,
respectively, the simulation results are still in excellent agreement with
the Riemann solutions. How to model deeper thermodynamic non-equilibrium flows by DB is indicated.
Via the DB method, it convenient to simulate nonequilibrium flows without knowing exact form of the hydrodynamic equations.
\end{abstract}

 \maketitle

\section{Introduction}

Compressible flows are ubiquitous in nature, industrial processes, weapons
physics, and even daily life. Typical examples are high-Mach-number
combustion system \cite{Combustion}, hydrodynamic instabilities in inertial confinement
fusion \cite{ICF}, strong shock waves driven by detonation \cite{SciChina}, etc. However, modeling and
simulation of such kind of fluids, especially containing shock waves and
contact discontinuities, is a challenging issue. The challenge arises from
the common and typical features of these fluids: (i) plenty of multi-scale
structures and interfaces, e.g., material and mechanical
interfaces (shock wave, rarefaction wave, detonation wave, etc); (ii)
competition between various spatio-temporal scales and kinetic modes; (iii)
complex forces and responses. Generally speaking, such a system is in a
global and/or local non-equilibrium state, which may challenge the validity
of the macroscopic models. Because the Euler equations assume that the
system is always in its local thermodynamic equilibrium. The Navier-Stokes
consider the thermodynamic non-equilibrium (TNE) via viscous stress
and heat flux which are quite dense or coarse-grained descriptions of TNE.
According to the Chapman-Enskog analysis \cite{Chapman}, the Navier-Stokes work only when the Knudsen number
is very small, in other words, only when the system deviates slightly from its thermodynamic equilibrium.
To access the more complicated and deeper non-equilibrium behaviors, one
possible strategy is to use the molecular dynamics (MD) or direct simulation
Monte Carlo method. But the spatial and temporal scales that the two schemes can access are
too small to be comparable with experiments.

As a bridge between the microscopic MD and macroscopic hydrodynamic
description, the mesoscopic Boltzmann equation (BE) is capable of handing
TNE effects for the full range of Knudsen number and Mach number \cite{Chapman}. However, due to the inherent nonlinearity and the
multiple integro-differential nature, direct solution of the full BE is
usually difficult. Thus, to develop approximate and simplified kinetic models
that preserve the most relevant features of the original BE has attracted extensive interest \cite{2004-2009-Li-JCP,2015-Li-PAS,2014-Zhang-JFM,2015-Zhang-JCP,Zhong-PRE2017,Zhong-POF2017}.
The discrete Boltzmann (DB) method \cite{2014-AG-PRE,2015-AG-PRE,2015-AG-SM,2016-AG-PRE,2016-AG-CF,2016-AG-CF2,Succi-JCP} which is developed
from the well-known, powerful, and promising lattice Boltzmann (LB) method \cite{Succi,EPL2038,Succi2017,2006-Shan-JFM,2013-Zhang-JFM,Sun2009,Tian2,Huang2015,Guo2016,
Liu2016,Liu2017,Chen2015,Zhong2012,Zhang-JCP2014,Wang2010,Yeomans,Qiu1,Qiu2} belongs to this category.
The BE presents values and evolutions of all kinetic moments of the distribution function.
Similar to, but different from, the original BE, the DB model presents not only values and evolutions of conserved kinetic moments (density, momentum and energy) but also those of some nonconserved kinetic moments. The former correspond to those described by hydrodynamic equations, the latter complement the former in finer description of the nonequilibrium flows \cite{InTech2018}.
In recent years, the DB model has brought some new physical insights into the fundamental mechanisms of various complex flow systems.
For example, the TNE intensity has been used to discriminate the spinodal decomposition stage and the domain growth stage in phase separation \cite{2015-AG-SM}; the abundant TNE characteristics have been used to distinguish and capture various interfaces \cite{2014-AG-PRE,2016-AG-PRE} in numerical experiments, to investigate the fundamental mechanisms for entropy increase \cite{2016-AG-CF2} in complex flows.
Some of the new observations brought by DB model, for example, the nonequilibrium fine structures of shock waves, have been confirmed and supplemented by the results of molecular dynamics \cite{Kang-1,Kang-2,Liu2017Recent}.

In practice, given the great importance of compressible TNE flows,
proposing compressible LB/DB models has attracted considerable attention since the early years of research.
For example, as early as 1992, Alexander \textit{et al.} \cite{Alexander} presented a
tunable sound speed model to soften sound speed and enhance Mach number.
Afterwards, Yan \textit{et al.}'s three-speed-three-energy-level model \cite{Yan},
Sun's adaptive model \cite{Sun1998}, Watari and Kataoka \textit{et al.}'s
multispeed models \cite{Kataoka-D2V16,Watari-PRE-2003,Watari-PRE-2004}, Qu \textit{et al.}'s circular function
model \cite{Qu2007}, Li \textit{et al.}'s double-distribution-function  (DDF)
model \cite{Li2007}, and multiple-relaxation-time model by our group \cite{MRT}, etc.,
appeared successively.

Compared to various 2D models, three-dimensional (3D) models for compressible Euler, NS and Burnett systems
with high speed are limited \cite{Sun2000-3D,Kataoka-PRE-2004-3D,Watari-PA-2006-3D,Li-2009-PLA-3D,Li-2013-PA-3D,Shu-5}. Among
these models, Sun \textit{et al.} \cite{Sun2000-3D} further extended their 2D locally adaptive
model to the 3D case in the framework of the traditional ``propagation +
collision" mode. The distinctive characteristic of the model is that the
particle velocities vary according to the local velocity and internal
energy, making it suitable for more extensive system with high Mach number.
After that, Kataoka \textit{et al.} \cite{Kataoka-PRE-2004-3D} contributed a highly efficient candidate at the Euler level with 15
discrete velocities.
Watari \textit{et al.} formulated 3D LB models for Euler and Navier-Stokes
equations (NSEs), respectively. The core idea is the application of discrete
velocity model (DVM) with larger number of velocities and higher isotropic
to introduce higher-order velocity terms in the discrete equilibrium
distribution function (DEDF) and to recover the correct target hydrodynamic
equations \cite{Kataoka-PRE-2004-3D,Watari-PA-2006-3D,Watari-PA-2007-3D}. This implementation advances the Mach number that can be stably
simulated to $2$.
The last two 3D LB models are both from He's group.
The first one utilizes a spherical function which satisfies required kinetic
moments to replace the Maxwellian distribution function (MDF) \cite{Li-2009-PLA-3D}, then $25$
discrete velocities are employed to discretize the simple function. The
maximum Mach number that the model can undergo is as high as $2.9$ in real
simulations.
To overcome the shortcomings: fixed specific heat
ratio and the lack of heat conduction effects in the energy equation, afterwards they proposed a coupling model which is
a combination of the DDF approach and the Hermite expansion approach for
constructing DEDF \cite{Li-2013-PA-3D}. The hybrid model is stable as
Mach number $Ma=10$.
Different from approaches mentioned above, recently, Shu and Lu \textit{et al.}
present a platform for constructing non-free parameter compressible LB models \cite{Shu-1,Shu-2,Shu-3,Lu,Shu-5,Shu-6}, from 1D to 3D, where both the DEDF and lattice velocities are automatically
determined from the needed and additional higher-order kinetic moments. Due to the lattice velocities are
determined physically without artificial selection, models proposed via this approach can be
applied to simulate compressible flows with a wide range of Mach numbers and complex configurations.

In this paper, we aim to further develop physically reasonable
and numerically efficient, multi-scale kinetic models to
meet the urgent requirements in engineering practice.
To this aim, we resort to the DB method \cite{InTech2018}.
Specifically, through inversely solving the kinetic moment relations that it should satisfy, we
propose a series of 3D DB models for compressible flows in and out of
equilibrium. Then, compare our approach for constructing DEDF with the common
approaches, such as the polynomial approach where DEDF is expanded in terms
of macroscopic variables with a global weighting coefficient or distribute
weighting coefficient; analyze their similarities, differences and
advantages. Finally, point out indications for designing
higher-order DB models, for example at the Burnett and super-Burnett levels.

\section{3D DB models at various levels}

In this section, a series of DB models at various levels are formulated.
To determine the departure levels from equilibrium state, we first deduce the
generalized hydrodynamic equations via the Chapman-Enskog procedure. The
starting point is the continuous Boltzmann equation with the BGK collision
term
\begin{equation}
\partial _{t}f+\mathbf{v}\cdot \bm{\nabla }f=-\frac{1}{\tau }[f-f^{(0)}]%
\mathtt{,}  \label{BE}
\end{equation}%
where $f$ represents the distribution function, $f^{(0)}=\rho (\frac{1}{2\pi
T})^{3/2}(\frac{1}{2n\pi T})^{1/2}\mathrm{exp}[-\frac{\mathbf{(v-u)}^{2}}{2T}%
-\frac{\eta ^{2}}{2nT}]$ is the 3D MDF, $\eta $ is a free parameter
describing the $n$ extra/internal degrees of freedom. Performing dynamically
the density, momentum and energy kinetic moments of Eq. (\ref{BE}), yields,
respectively, the continuity, momentum conservation and energy conservation
equations
\begin{equation}
\partial _{t}\rho +\bm{\nabla }\cdot (\rho \mathbf{u})=0\text{,}  \label{NS1}
\end{equation}%
\begin{equation}
\partial _{t}(\rho \mathbf{u)}+\bm{\nabla }\cdot (\rho \mathbf{uu+}P\mathbf{%
I+\Delta }_{2}^{\ast })=0\text{,}  \label{NS2}
\end{equation}%
\begin{equation}
\partial _{t}[\rho (e+\frac{\mathbf{u}^{2}}{2})]+\bm{\nabla }\cdot \lbrack
\rho \mathbf{u}(e+T+\frac{\mathbf{u}^{2}}{2})+\mathbf{\Delta }_{2}^{\ast
}\cdot \mathbf{u+\Delta }_{3,1}^{\ast }]=0\text{,}  \label{NS3}
\end{equation}%
where $P=\rho T$ and $e=\frac{n+3}{2}T$ stand for pressure and specific
internal energy, respectively. $\mathbf{\Delta }_{2}^{\ast }$ ($\mathbf{%
\Delta }_{3,1}^{\ast }$) is the non-organised moment fluxes (non-organised
energy fluxes), presented by our group and defined as \cite{2016-AG-CF2}
\begin{equation}
\mathbf{\Delta }_{2}^{\ast }=\mathbf{M}_{2}^{\ast }(f-f^{(0)})=\iint
(f^{(1)}+f^{(2)}+...)\mathbf{v}^{\ast }\mathbf{v}^{\ast }d\mathbf{v}d\eta
\text{,}  \label{D2}
\end{equation}%
\begin{equation}
\mathbf{\Delta }_{3,1}^{\ast }=\mathbf{M}_{3,1}^{\ast }(f-f^{(0)})=\iint
(f^{(1)}+f^{(2)}+...)\frac{\mathbf{v}_{i}^{\ast 2}+\eta _{i}^{2}}{2}\mathbf{v%
}^{\ast }d\mathbf{v}d\eta \text{,}  \label{D31}
\end{equation}%
where $f^{(i)}$ represents the $i$-th order derivation from $f^{(0)}$, $%
\mathbf{v}^{\ast }=\mathbf{v-u}$ the thermal velocity, $\mathbf{M}_{2}^{\ast
}$ and $\mathbf{M}_{3,1}^{\ast }$ are kinetic central moments. It is well
known that the Euler equations assume that the system is always in its local
thermodynamic equilibrium (mechanical non-equilibrium) $f=f^{(0)}$. So both
of the two TNE measures
\begin{equation}
\mathbf{\Delta }_{2}^{\ast }=0\text{,}
\end{equation}%
\begin{equation}
\mathbf{\Delta }_{3,1}^{\ast }=0\text{.}
\end{equation}%
The NSEs only consider TNE effects induced by $f^{(1)}$ through the quite
dense quantities, i.e., the viscous stress and heat flux. Therefore,
\begin{equation}
\mathbf{\Delta }_{2}^{\ast }=\iint f^{(1)}\mathbf{v}^{\ast }\mathbf{v}^{\ast
}d\mathbf{v}d\eta =-\mu \lbrack \bm{\nabla}\mathbf{u}+(\bm{\nabla}\mathbf{u}%
)^{T}-\frac{2}{n+3}\mathbf{I}\bm{ \nabla }\cdot \mathbf{u}]=-\bm{\sigma}_{%
\text{NS}}\text{,}
\end{equation}%
\begin{equation}
\mathbf{\Delta }_{3,1}^{\ast }=\iint \frac{1}{2}f^{(1)}(\mathbf{v}_{i}^{\ast
2}+\eta _{i}^{2})\mathbf{v}^{\ast }d\mathbf{v}d\eta =-\kappa \bm{\nabla }T=
-\mathbf{j}_{q-\text{NS}}\text{,}
\end{equation}%
where $\mu =\rho T\tau $ and $\kappa =c_{p}\rho T\tau $ are viscosity
coefficient and heat conductivity coefficient, $c_{p}=\frac{5+n}{2}$ is the
specific-heat at constant pressure. Similarly, the Burnett equations
consider the TNE effects induced by $f^{(1)}+f^{(2)}$. As a result,
\begin{equation}
\mathbf{\Delta }_{2}^{\ast }=\iint (f^{(1)}+f^{(2)})\mathbf{v}^{\ast }%
\mathbf{v}^{\ast }d\mathbf{v}d\eta =-\bm{\sigma}_{\text{Burnett}}\text{,}
\end{equation}%
\begin{equation}
\mathbf{\Delta }_{3,1}^{\ast }=\iint \frac{1}{2}(f^{(1)}+f^{(2)})(\mathbf{v}%
_{i}^{\ast 2}+\eta _{i}^{2})\mathbf{v}^{\ast }d\mathbf{v}d\eta =-\mathbf{j}%
_{q-\text{Burnett}}\text{,}
\end{equation}%
where $\bm{\sigma}_{\text{Burnett}}$ and $\mathbf{j}_{q-\text{Burnett}}$ are
the Burnett viscous stress and heat flux which are primarily unknown. For
the super-Burnett equations, TNE effects contributed from $f^{(3)}$ should
be further considered.

Here it is necessary to point out that, the DB method presents two sets of measures for the
TNE. One set is dynamically from the difference of $f$ and $f^{(0)}$,
i.e., $\mathbf{\Delta }_{m,n}^{\ast }$, such as $\mathbf{\Delta }_{2}^{\ast
} $, $\mathbf{\Delta }_{3,1}^{\ast }$, $\mathbf{\Delta }_{3}^{\ast }$ and $%
\mathbf{\Delta }_{4,2}^{\ast }$. The other set includes the viscous stress
and heat flux discussed above. The former one describes the specific TNE status,
the latter one describes the influence of those TNE to the macroscopic control
equations. The former one is local, the latter one is non-local. The former one
is finer, the latter one is coarser. Compared to macroscopic hydrodynamic
descriptions, the TNE manifestations $\mathbf{\Delta }_{m,n}^{\ast }$
present a simple, convenient and effective measure to the departure of the
system from the local thermodynamic equilibrium.

To conduct numerical simulations, the formost important step is to construct
DEDF $f_{i}^{(0)}$. In detail, choosing an appropriate DVM $\mathbf{v}_{i}$
to discretize $f^{(0)}$, or, in other words, to remove the continuous
dependency on the velocity space, where $i=1,2,3,...,N$ and $N$ is the
number of velocities in DVM. Chapman-Enskog multi-scale analysis tells us
that, to recover different targeted equations, $f_{i}^{(0)}$ should satisfy
different kinetic moment relations. Specifically, the following five are
necessary to satisfy to recover the hydrodynamic equations at the Euler
level
\begin{equation}
\mathbf{M}_{0}=\sum_{i}f_{i}^{(0)}=\rho ,  \label{M0}
\end{equation}%
\begin{equation}
\mathbf{M}_{1}=\sum_{i}f_{i}^{(0)}\mathbf{v}_{i}=\rho \mathbf{u}\text{,}
\label{M1}
\end{equation}%
\begin{equation}
\mathbf{M}_{2,0}=\sum_{i}f_{i}^{(0)}(\mathbf{v}_{i}^{2}+\eta _{i}^{2})=\rho
\lbrack (n+3)T+\mathbf{u}^{2}\mathbf{]},  \label{M20}
\end{equation}%
\begin{equation}
\mathbf{M}_{2}=\sum_{i}f_{i}^{(0)}\mathbf{v}_{i}\mathbf{v}_{i}=\rho (T%
\mathbf{I}+\mathbf{uu})\text{,}  \label{M2}
\end{equation}%
\begin{equation}
\mathbf{M}_{3,1}=\sum_{i}f_{i}^{(0)}(\mathbf{v}_{i}^{2}+\eta _{i}^{2})%
\mathbf{v}_{i}=\rho \mathbf{u}[(n+5)T+\mathbf{u}^{2}]\text{,}  \label{M31}
\end{equation}%
where $\mathbf{I}$ represents a unit matrix. To recover the
hydrodynamic equations at the NS level, the following two are also required
\begin{equation}
\mathbf{M}_{3}=\sum_{i}f_{i}^{(0)}\mathbf{v}_{i}\mathbf{v}_{i}\mathbf{v}%
_{i}=\rho \lbrack T(\mathbf{u}_{\alpha }\mathbf{e}_{\beta }\mathbf{e}%
_{\gamma }\delta _{\beta \gamma }+\mathbf{e}_{\alpha }\mathbf{u}_{\beta }%
\mathbf{e}_{\gamma }\delta _{\alpha \gamma }+\mathbf{e}_{\alpha }\mathbf{e}%
_{\beta }\mathbf{u}_{\gamma }\delta _{\alpha \beta })+\mathbf{uuu}]\text{,}
\label{M3}
\end{equation}%
\begin{equation}
\mathbf{M}_{4,2}=\sum_{i}f_{i}^{(0)}(\mathbf{v}_{i}^{2}+\eta _{i}^{2})%
\mathbf{v}_{i}\mathbf{v}_{i}=\rho \lbrack (n+5)T^{2}+T\mathbf{u}^{2}\mathbf{%
]I}+\rho \lbrack (n+7)T+\mathbf{u}^{2}]\mathbf{uu}\text{.}  \label{M42}
\end{equation}%
Moreover, to recover the hydrodynamic equations at the Burnett (super-Burnett) level, kinetic moments $
\mathbf{M}_{4}$ ($\mathbf{M}_{5}$) and $\mathbf{M}_{5,3}$ ($\mathbf{M}_{6,4}$%
) are further required.

From analysis above, it can be seen that, the physically concerned hydrodynamic
quantities ($\rho $, $\mathbf{u}$, $P$, $T$, $\bm{\sigma}$, $\mathbf{j}_{q}$%
, etc) are some kinetic moments of the distribution function $f$. These conserved and nonconserved
quantities can
finally be roughly calculated from some kinetic moments of the local
equilibrium distribution function $f^{(0)}$. The calculation of any
non-conserved quantity triggers the requirement of higher-order kinetic
moments of $f^{(0)}$. When constructing the DB model, it must be ensured that the
required kinetic moments of $f^{(0)}$, originally in integral form, equal to,
those in the discrete summation form,
\begin{equation}
\mathbf{M}_{m,n}=\iint f^{(0)}(\mathbf{v}^{2}+\eta ^{2})^{\frac{m-n}{2}}%
\mathbf{v}^{n}d\mathbf{v}d\eta =\sum_{i}f_{i}^{(0)}(\mathbf{v}_{i}^{2}+\eta
_{i}^{2})^{\frac{m-n}{2}}\mathbf{v}_{i}^{n}\text{.}
\end{equation}

Subsequently, we present an effective approach for constructing DEDF.
Equations (\ref{M0})-(\ref{M42}) can be rewritten in a matrix form
\begin{equation}
\mathbf{M=C\times f}^{(0)}\text{,}  \label{Feq_1}
\end{equation}%
where $\mathbf{M}=(M_{0},M_{1x},M_{1y},M_{1z},...,M_{4,2zz})^{T}$ is a $%
30\times 1$ matrix, with each element represents the moment of $f_{i}^{(0)}$. $\mathbf{C%
}=(\mathbf{C}_{1}\mathbf{,C}_{2}\mathbf{,\cdots ,C}_{30})$ is a $30\times 30$
square matrix bridging DEDF and the kinetic moments with $\mathbf{C}%
_{i}=(1,v_{ix},v_{iy},v_{iz}...,(\mathbf{v}_{i}^{2}+\eta
_{i}^{2})v_{iz}v_{iz})^{T}$, $\mathbf{f}^{(0)}=(f_{1}^{(0)},f_{2}^{(0)},%
\cdots ,f_{30}^{(0)})^{T}$. Therefore, we obtain $\mathbf{f}^{(0)}$ through
the following way
\begin{equation}
\mathbf{f}^{(0)}={{\mathbf{C}}^{-1}}\mathbf{\times M}\text{,}  \label{Invm}
\end{equation}%
with $\mathbf{C}^{-1}$ the inverse of $\mathbf{C}$. The next step is to design DVM to
ensure the existence of $\mathbf{C}^{-1}$. For the level of Euler (NS)
equations, a 3D DVM with $14$ ($30$) discrete velocities is appropriate, as
shown in Table I. For the level of Burnett equations, a DVM with $55$
discrete velocities is necessary and enough. It is clear that, the D3V30 DVM
contains three parts: (i) the D3V14 model, (ii) an additional symmetric
D3V12 model, and (iii) an antisymmetric part D3V4. The former two symmetric
parts help to improve the spatial isotropy, the antisymmetric part together
with $\eta _{i}$ aims to guarantee the existence of $\mathbf{C}^{-1}$. So
the choosing of the DVM is very flexible. Here for the D3V14 model, we set $\eta _{i}=\eta
_{0}$ when $i$ is an odd number, otherwise $\eta _{i}=0$. For the D3V30
model, we set $\eta _{i}=\eta _{0}$ when $1\leq i\leq 6$, and otherwise $%
\eta _{i}=0$. As a result, the third part can be chosen as
\begin{equation}
\mathbf{C}_{\text{D3V4}}=c\left(
\begin{array}{cccc}
-2 & -2 & 1 & 3 \\
1 & 3 & -2 & -2 \\
2 & -2 & -1 & 1%
\end{array}%
\right) \text{; }c\left(
\begin{array}{cccc}
-1 & -2 & 2 & 1 \\
1 & 2 & -1 & -2 \\
2 & -2 & -1 & 1%
\end{array}%
\right) \text{; }c\left(
\begin{array}{cccc}
2 & -2 & -1 & 1 \\
1 & 2 & -1 & -2 \\
-1 & -2 & 2 & 1%
\end{array}%
\right)
\end{equation}%
Here $c$ and $\eta _{0}$ are two free parameters, adjusted to optimize the
properties of the model.
\begin{table}[tbp]
\begin{tabular}{c||c}
\hline\hline
DVM & $\left(
\begin{array}{ccc}
v_{1x},v_{2x},\dots ,v_{Nx} &  &  \\
v_{1y},v_{2y},\dots ,v_{Ny} &  &  \\
v_{1z},v_{2z},\dots ,v_{Nz} &  &
\end{array}%
\right) $ \\ \hline
D3V14 & $c \left(
\begin{array}{cccccccccccccc}
1 & -1 & 0 & 0 & 0 & 0 & 1 & -1 & -1 & 1 & 1 & -1 & 1 & -1 \\
0 & 0 & 1 & -1 & 0 & 0 & 1 & -1 & 1 & -1 & -1 & 1 & 1 & -1 \\
0 & 0 & 0 & 0 & 1 & -1 & 1 & -1 & 1 & -1 & 1 & -1 & -1 & 1%
\end{array}%
\right) $ \\ \hline
D3V30 & $%
\begin{array}{ccccc}
\text{D3V14} & c \left(
\begin{array}{cccccccccccc}
0 & 0 & 0 & 0 & 1 & -1 & -1 & 1 & 1 & -1 & -1 & 1 \\
1 & -1 & -1 & 1 & 0 & 0 & 0 & 0 & 1 & -1 & 1 & -1 \\
1 & -1 & 1 & -1 & 1 & -1 & 1 & -1 & 0 & 0 & 0 & 1%
\end{array}%
\right) & \text{D3V4} &  &  \\
\text{symmetric part I} & \text{symmetric part II} & \text{antisymmetric part%
} &  &
\end{array}%
$ \\ \hline\hline
\end{tabular}%
\caption{Discrete velocities in D3V14 and D3V30.}
\end{table}
The calculation of $\mathbf{C}^{-1}$ is a tough work, so we resort to
software, Mathematica; and here only the $\mathbf{C}_{\text{D3V14}%
}^{-1}$ for D3V14 model is exhibited for simplicity,
\begin{equation}
\mathbf{C}_{\text{D3V14}}^{-1}=\left(
\begin{array}{cccccccccccccc}
\frac{1}{4} & A & 0 & 0 & 2D & 0 & 0 & 0 & 0 & 0 & -D & -2F & 0 & 0 \\
\frac{1}{4} & -A & 0 & 0 & 0 & -2D & -2D & 0 & 0 & 0 & D & 2F & 0 & 0 \\
\frac{1}{4} & 0 & A & 0 & 0 & 2D & 0 & 0 & 0 & 0 & -D & 0 & -2F & 0 \\
\frac{1}{4} & 0 & -A & 0 & -2D & 0 & -2D & 0 & 0 & 0 & D & 0 & 2F & 0 \\
\frac{1}{4} & 0 & 0 & A & -2D & -2D & 0 & 0 & 0 & 0 & D & 0 & 0 & -2F \\
\frac{1}{4} & 0 & 0 & -A & 0 & 0 & 2D & 0 & 0 & 0 & -D & 0 & 0 & 2F \\
0 & -B & -B & 0 & -L & -L & -L & 0 & D & D & L & F & F & 0 \\
-\frac{1}{8} & B & B & 0 & J & J & J & D & 0 & 0 & -L & -F & -F & 0 \\
-\frac{1}{8} & B & 0 & -B & J & J & J & 0 & -D & 0 & -L & -F & 0 & F \\
0 & -B & 0 & B & -L & -L & -L & -D & 0 & D & L & F & 0 & -F \\
-\frac{1}{8} & 0 & B & -B & J & J & J & 0 & 0 & -D & -L & 0 & -F & F \\
0 & 0 & -B & B & -L & -L & -L & -D & D & 0 & L & 0 & F & -F \\
-\frac{1}{8} & 0 & 0 & 0 & H & H & H & D & -D & -D & -G & 0 & 0 & 0 \\
0 & 0 & 0 & 0 & -G & -G & -G & 0 & 0 & 0 & G & 0 & 0 & 0 \\
\end{array}%
\right) \text{,}
\end{equation}%
with $A=\frac{3}{4c}$, $B=\frac{1}{8c}$, $D=\frac{1}{4c^{2}}$, $F=\frac{1}{%
8c^{3}}$, $G=\frac{1}{\eta _{0}^{2}}$, $H=\frac{1}{8c^{2}}+\frac{1}{\eta
_{0}^{2}}$, $J=\frac{3}{8c^{2}}+\frac{1}{\eta _{0}^{2}}$, $L=\frac{1}{4c^{2}}%
+\frac{1}{\eta _{0}^{2}}$.

It is necessary to further point out that, the choosing of $c$ and $\eta_i$
is artificial and empirical. Nevertheless,
fortunately, practical simulations indicate that the stable parameter range
for models proposed by this approach is fairly wide.
For example, for a weak shock tube problem, e.g., the Sod shock tube, the stable
range is $\frac{1}{2}u_{max}<c<2u_{max}$ and $\eta >\frac{n+3}{4}u_{max}$;
while for a strong shock tube problem, e.g., the collision of two string shocks,
the stable range shrinks to $\frac{1}{2}u_{max}<c<u_{max}$ and $\eta >\frac{%
n+3}{4}u_{max}$.

\section{Comparisons with the common approach for constructing DEDF}

In this subsection, our approach for constructing DEDF is compared with the
commonly adopted polynomial approach where DEDF is expanded in terms of
macroscopic quantities. It is noted that, the similarities of the two approaches are
the solution of the required kinetic moments, the adoption of DVM with less number of velocities.
In contrast, the present approach is (i) conceptually
simple, concise and straightforward, (ii) physically flexible and general,
and (iii) numerically efficient and stable. Subsequently, we would like to
illustrate these advantages one by one.

\subsection{conceptually simple, concise and straightforward}

In the common approach, the DEDF $f_{i}^{(0)}$ is conveniently expanded in
terms of the macroscopic variables ($\rho $, $\mathbf{u}$, and $T$) by only
keeping the first relevant orders in the expansion. In detail, according to
the nice work of Watari \emph{et al.} \cite{Watari-PRE-2004}, there exist
two ways to formulate DEDF. One is to distribute weighting coefficient on
the whole expansion equation, i.e, the DEFD owns a global weighting
coefficient. For convenience of description, we refer to this approach as
the global way. For example, to recover the 2D NSEs in the continuum limit,
it is reasonable to expand the local DEDF as the polynomial of the flow
velocity up to the fourth order from the MDF
\begin{eqnarray}
f_{ki}^{(0)} &=&\frac{\rho }{2\pi T}\exp [-\frac{(\mathbf{v}_{ki}-\mathbf{u}%
)^{2}}{2T}]=\frac{\rho }{2\pi T}\exp [-\frac{\mathbf{v}_{k}^{2}}{2T}]\exp [%
\frac{\mathbf{v}_{ki}-\mathbf{u}^{2}/2}{T}]  \notag \\
&=&\rho F_{k}[(1-\frac{\mathbf{u}^{2}}{2T}+\frac{\mathbf{u}^{4}}{8T^{2}})+%
\frac{\mathbf{v}_{ki}\cdot \mathbf{u}}{T}(1-\frac{\mathbf{u}^{2}}{2T})+\frac{(%
\mathbf{v}_{ki}\cdot \mathbf{u})^{2}}{2T^{2}}(1-\frac{\mathbf{u}^{2}}{2T})  \notag
\\
&&+\frac{(\mathbf{v}_{ki} \cdot \mathbf{u})^{3}}{6T^{3}}+\frac{(\mathbf{v}_{ki} \cdot
\mathbf{u})^{4}}{24T^{4}}],  \label{feq}
\end{eqnarray}%
where $F_{k}=\frac{1}{2\pi T}\exp [-\frac{\mathbf{v}_{k}^{2}}{2T}]$ is the
global weighting coefficient. Examples for compressible flows which fall into
this category are referred to the works of Watari \emph{et al.} \cite
{Watari-PRE-2003,Watari-PA-2006-3D,Watari-PA-2007-3D}, Sofonea
\emph{et al.} \cite%
{Sofonea-PRE-2005,Sofonea-EPL-2006,Sofonea-PRE-2006,Sofonea-PRE-2007,Sofonea-JCP-2009}
and our group \cite%
{Xu-EPL-2005,Xu-PRE-2005,Xu-PA-2008,Xu-PRE-2011-1,Xu-PRE-2011-2,Xu-EPL-2012}, etc.

To calculate the DEDF numerically, one needs to determine the global
factor $F_{k}$ at first. It should be noted that $F_{k}$ cannot be calculated
directly from its definition, while it should take values in such a way that
satisfies the required kinetic moment relations. Essentially, these
constraints on DEDF are transferred to the global coefficient $F_{k}$.
Generally, it is necessary to take the following steps for deriving the
analytical formulation of $F_{k}$: (i) substituting Eq.(\ref{feq}) into Eqs.(%
\ref{M0})-(\ref{M42}) and comparing the terms with the same orders of $%
\mathbf{u}$; (ii) simplifying and obtaining requirements on $F_{k}$; (iii)
considering the isotropic properties of DVM and finding the least
requirements on $F_{k}$; (iv) choosing approximate velocity speeds $v_{k}$
and analytical giving the specific formulation of $F_{k}$.
So it is a tedious and challenging procedure to determine the DEDF by
this way.

The second way is to distribute weighting coefficients to each power term of
the expansion equation. For the convenience of description, we refer to this way
as the distributed way. For instance, we also aim to recover the NSEs, then
the local DEDF can be expanded as follows \cite{Kataoka-D2V16}:
\begin{eqnarray}
f_{i}^{(0)} &=&\rho \lbrack a_{0i}+a_{1i}T+a_{2i}T^{2}+(a_{3i}+a_{4i}T)%
\mathbf{u}^{2}+a_{5i}\mathbf{u}^{4}+(b_{0i}+b_{1i}T+b_{2i}\mathbf{u}^{2})(%
\mathbf{v}_{i}\cdot \mathbf{u})  \notag \\
&&+(d_{0i}+d_{1i}T+d_{2i}\mathbf{u}^{2})(\mathbf{v}_{i}\cdot \mathbf{u}%
)^{2}+e_{i}(\mathbf{v}_{i}\cdot \mathbf{u})^{3}],  \label{feq2}
\end{eqnarray}%
where $a_{0i}$, $a_{1i}$, $a_{2i}$, $a_{3i}$, $a_{4i}$, $a_{5i}$, $b_{0i}$, $%
b_{1i}$, $b_{2i}$, $d_{0i}$, $d_{1i}$, $d_{2i}$ and $e_{i}$ are distributed
weighting coefficients. Examples for compressible flows that fall into this
category are referred to the works of Alexander \emph{et al.} \cite
{Alexander-PRE-1993}, Chen \emph{et al.} \cite{Chen-PRE-1994}, Yan \emph{et
al.} \cite{Yan}, Kataoka \emph{et al.} \cite%
{Kataoka-D2V16,Kataoka-PRE-2004-3D}, Watari \emph{%
et al.} \cite{Watari-PRE-2004} and our group \cite{Xu-IJMPC-2007,Xu-CTP-2009,3D-CTP}, etc. The deriving process of $a_{0i}$%
,...,$e_{i}$ is identical to what we described in the first way. Also, it is
a cumbersome and difficult work to give the specific formulations of so many
free parameters in Eq. (\ref{feq2}).

In the present approach, the only one step needed is to inversely calculate the DEDF
via the way described in Eq. (\ref{Invm}). Therefore, compared to the usual
approaches reported above, our methodology seems more conceptually simple, concise
and straightforward. With the help of software the deriving process is nearly automatic.

\subsection{physically flexible and general}

It is interesting to find that, the deriving process of DEDF is totally
associate with the choosing of DVM. Only if the DVM has enough isotropy, the
hydrodynamic equations can be recovered from the DB equation successfully.
From the works of Watari and Kataoka above, we conclude that: (i) the
choosing of DVM is a complicated process, it should carefully consider the
direction and symmetry of the DVM, i.e., the isotropy of DVM, so as to
recover the target hydrodynamic equations accurately; it should carefully
take the least number and sizes of the discrete velocities into account so
as to improve the computational efficiency and ensure stability of the
model; (ii) the distributed way is more economical than the global way.
Specifically, Kataoka \emph{et al.} \cite{Kataoka-D2V16} use a D2V16
model to recover the NSEs. To the best of our knowledge, this is the DVM
with the least number of discrete velocities that can correctly recover the
hydrodynamic equations at the NS level. Nevertheless, Watari \emph{et al.}
\cite{Watari-PA-2007-3D} employe a D2V65 model to recover the same equations.
Undoubtedly, it increases the computational load and complexity. In contrast, the choosing of DVM in our approach is more flexible. In
the present framework, only if we ensure $|\mathbf{C}^{-1}|\neq 0$, then the
DVM is sufficient to recover the aimed macroscopic equations.

Moreover, the present approach is more general, it works consistently for
constructing DB models of different dimensions (1D, 2D, and 3D) \cite{2D} and at
various levels (Euler, NS, Burnett, and super-Burnett).

\subsection{numerically efficient and stable}

It is noteworthy that, in our approach, the number of discrete velocities in
the DVM equals to that of the kinetic moment relations. Thus, our approach adapts the minimal set of velocities in
the phase space, resulting in the highest computational efficiency.
Particularly, $30$ discrete velocities is enough for recovering the NSEs by
using our approach, while $73$ discrete velocities are needed to recover the
Euler equations through the global way.

Most of the LB models schemed through the polynomial way are only suitable
for compressible flows with small Mach number, except for those in Refs.
\cite{Xu-PA-2008,Xu-IJMPC-2007,3D-CTP,Watari-PA-2007-3D}%
. The dominate reasons are as below. First, owing to the DEDF is based on
a low-Mach-number Taylor expansion of the MDF, thus it cannot describe compressible flows with high speed
because of the insufficient truncation in the DEDF and the
insufficient isotropy in the DVM. Obviously, the terms with higher
orders of velocity contribute more than those with lower orders when the
velocity is larger than $1$. In our approach, the ``insufficient
truncation" is avoidable.

Secondly, we find that in some models belonging to the distributed way \cite{Kataoka-D2V16}, the sizes of DVM are fixed; but in the
global way, they are free. Numerical simulations and stability analysis \cite{Xu-PA-2008} demonstrate that the sizes of DVM have
significant effects on the stability of LB models. The freedom in the sizes
of DVM contributes to enhance the stability. For example, the model in Ref. \cite{Watari-PA-2007-3D} can be used to simulate subsonic and/or supersonic flows.
Also in our work, the sizes of DVM vary with the specific test we studied.
Consequently, through adjusting the sizes of the discrete velocities the
stably simulated Mach number can remarkably increase up to $100$ or
even higher.

\section{Numerical tests and analysis}

In this section, several typical benchmarks, ranging from 1D to 3D, from
subsonic to hypersonic, are adopted to examine the models. To improve the
numerical accuracy, stability, as well as efficiency, the third-order
implicit-explicit Runge-Kutta finite difference scheme \cite{IMEX} and the
second-order non-oscillatory non-free-parameter and dissipative finite
difference scheme \cite{NND} are utilized to discrete the temporal and
spatial derivatives, respectively.

\subsection{Couette flow}
The Couette flow is a classical test case which is employed to validate
the ability of the new model for describing viscous heat conducting flows
with various Mach numbers and flexible specific-heat ratios \cite{Shu-PRE2016-Couette}.
For this problem considered, a viscous fluid flow between two infinite parallel flat plates has the
initial conditions $(\rho ,T,u_{x},u_{y},u_{z})=(1.0,1.0,0.0,0.0,0.0)$. When
simulation starts, the top plate moves along the horizontal direction with
a speed $u_{0}=1.3$, while the bottom plate is stationary. The distance
between the plates is $H$. Periodic boundary conditions (PBCs) are used
in the $x$ and $z$-directions. For boundary nodes in the $y$-direction, the
nonequilibrium extrapolation method is applied. We carry out simulations
with $\Delta x=\Delta y=\Delta z=10^{-3}$, $\tau $ $=5\times
10^{-4}$, $N_{x}\times N_{y}\times N_{z}=4\times 65\times 4.$
Figure 1(a) shows comparisons of the DB simulation results and exact
solutions for the vertical velocity distributions at times $t=0.05$, $0.15$,
$0.40$, $1.00$, $20.0$.
It is clear that the simulation results agree excellently well with following analytical
solution
\begin{equation}
u=\frac{y}{H}u_{0}+\frac{2}{\pi }u_{0}\overset{\infty }{\underset{n=1}{\sum }%
}[\frac{(-1)^{n}}{n}\exp (-n^{2}\pi ^{2}\frac{\mu t}{\rho H^{2}})\sin (\frac{%
n\pi y}{H})]\text{.}
\end{equation}
Figure 1(b) plots temperature profiles along the $y$-direction in
steady Couette flow for cases with various specific-heat ratios $\gamma =5/3$%
, $7/5$, $4/3$.
The theoretical temperature distribution at the steady state is
\begin{equation}
T=T_{0}+\frac{\Pr}{2c_{p}}u_{0}^{2}\frac{y}{H}(1-\frac{y}{H})\text{,}
\end{equation}%
where $T_{0}$ is the temperature of the top/bottom wall.
The simulation results also match well with the analytical ones, even when the Mach number of the flow is
$Ma=1.12$.

\begin{figure}[tbp]
{%
\centerline{\epsfig{file=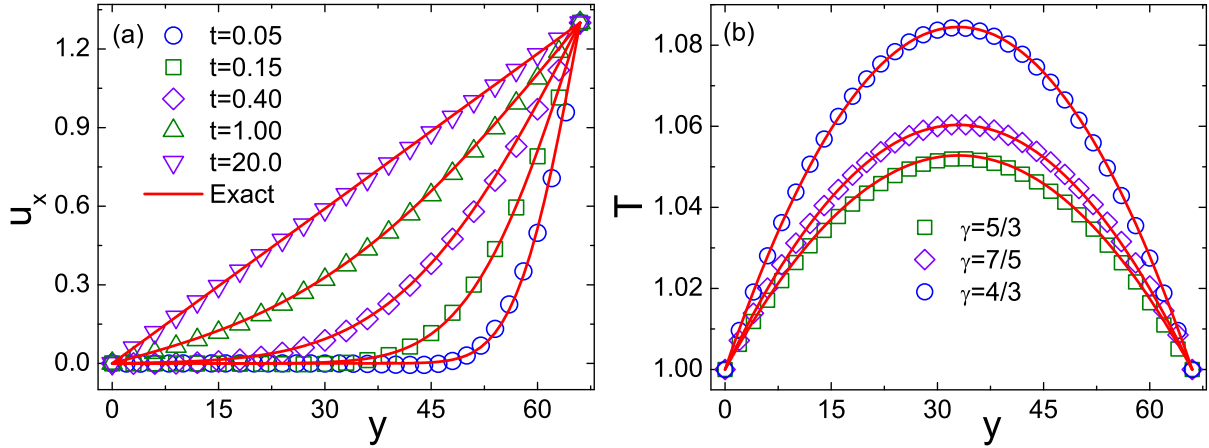,bbllx=1pt,bblly=0pt,bburx=590pt,bbury=266pt,
width=1.0\textwidth,clip=}}}
\caption{(a) velocity profiles in unsteady Couette flow at characteristic
times. (b) temperature profiles in steady Couette flow for cases with
various specific-heat ratios.}
\end{figure}

\subsection{1D Riemann problems}

\subsubsection{Sod shock tube}

\begin{figure}[tbp]
{%
\centerline{\epsfig{file=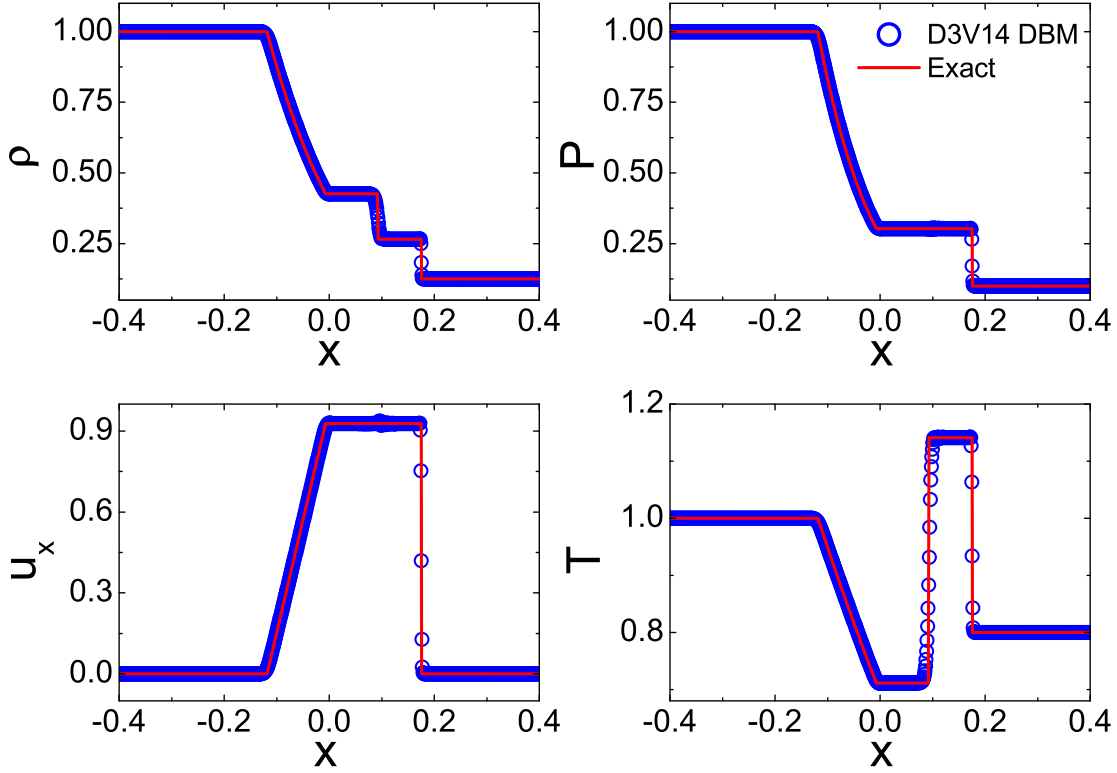,bbllx=1pt,bblly=8pt,bburx=537pt,bbury=388pt,
width=0.9\textwidth,clip=}}}
\caption{Comparisons between DB simulations and the exact solutions for the
Sod shock tube.}
\end{figure}

The Sod shock tube problem, introduced by Gary A. Sod in 1978, is a
classical and ideal test to verify the performance of models for
compressible flow. Analytical solution of this problem contains a
right-propagating shock wave, a left-propagating rarefaction wave, and a
contact discontinuity representing the interface. The initial conditions are
described by
\begin{equation*}
\left\{
\begin{array}{c}
(\rho ,T,u_{x},u_{y}, u_z)|_{L}=(1.0,1.0,0.0,0.0,0.0), \\
(\rho ,T,u_{x},u_{y}, u_z)|_{R}=(0.125,0.8,0.0,0.0,0.0),%
\end{array}%
\right.
\end{equation*}%
where subscripts ``L" (``R") indicates quantities at the left (right) side
of the discontinuity. In the $y$ and $z$ directions, we apply the PBCs.
In the $x$ direction, we impose the supersonic
boundary conditions: $f_{i,-1}=f_{i,0}=f_{i,1}^{(0)}$ on the left side, and $%
f_{i,N_x+2}=f_{i,N_x+1}=f_{i,N_x}^{(0)}$ on the right side, where $-1$, $0$,
$N_x+1$ and $N_x+2$ are the indexes of ghost nodes. In the computational
domain, a mesh $N_{x}\times N_{y}\times N_z=2000\times 2 \times 2$ with grid
sizes $\Delta x=\Delta y=\Delta z= 10^{-3}$. Other parameters are $%
\Delta t=10 \tau =10^{-4}$, $c=1.5$, $\eta _{0}=1.4$, and $\gamma =1.4$.
Figure 2 shows the comparisons between simulated results obtained from the
D3V14 DB model and Riemann solutions at $t=0.1$. The two sets of results
have a satisfying agreement. The characteristic structures, such as the
shock wave, the expanding wave and contact discontinuity are well captured
without spurious numerical oscillations.

\subsubsection{Lax shock tube}

\begin{figure}[tbp]
{%
\centerline{\epsfig{file=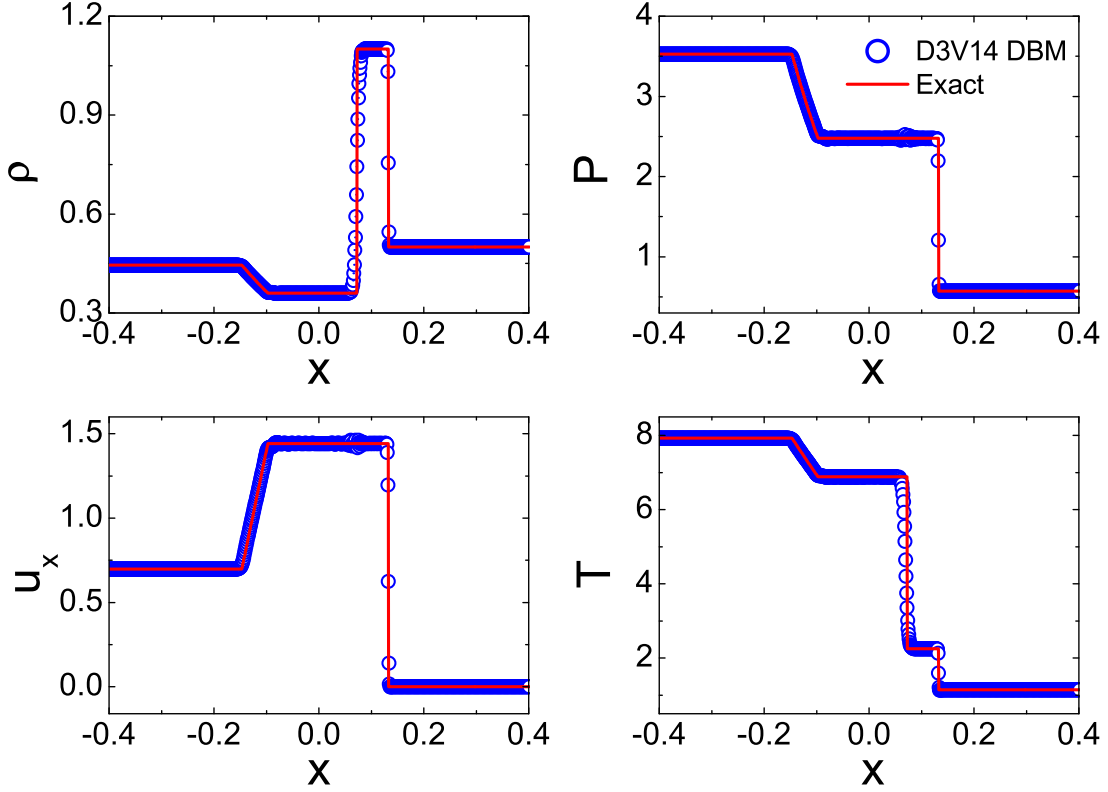,bbllx=8pt,bblly=2pt,bburx=578pt,bbury=413pt,
width=0.9\textwidth,clip=}}}
\caption{Comparisons between DB simulations and the exact solutions for the
Lax shock tube.}
\end{figure}

The second test is the Lax shock tube, physically similar to the Sod shock
tube problem, except for a discontinuity in the velocity along the $x$
direction and a sharper discontinuity in temperature. Concretely, the initial conditions are
\begin{equation*}
\left\{
\begin{array}{c}
(\rho ,T,u_{x},u_{y}, u_z)|_{L}=(0.445,7.928,0.698,0.0,0.0), \\
(\rho ,T,u_{x},u_{y}, u_z)|_{R}=(0.50,1.142,0.0,0.0,0.0).%
\end{array}%
\right.
\end{equation*}
The computed density, pressure, velocity, temperature and the corresponding
exact solutions at $t=0.05$ are plotted in Fig. 3. Here $c=2$, $\eta_{0}=3$,
$\tau=3 \times 10^{-5}$, $\gamma=5/3$, other parameters are unchanged. The
two sets of results also agree excellently with each other. Additionally,
the shock wave is captured sharply, without excessive numerical
dissipations. Enlargement of the part containing shock wave manifests that
the shock wave only spreads over three to four grid cells.

\subsubsection{Collision of two strong shocks}

\begin{figure}[tbp]
{%
\centerline{\epsfig{file=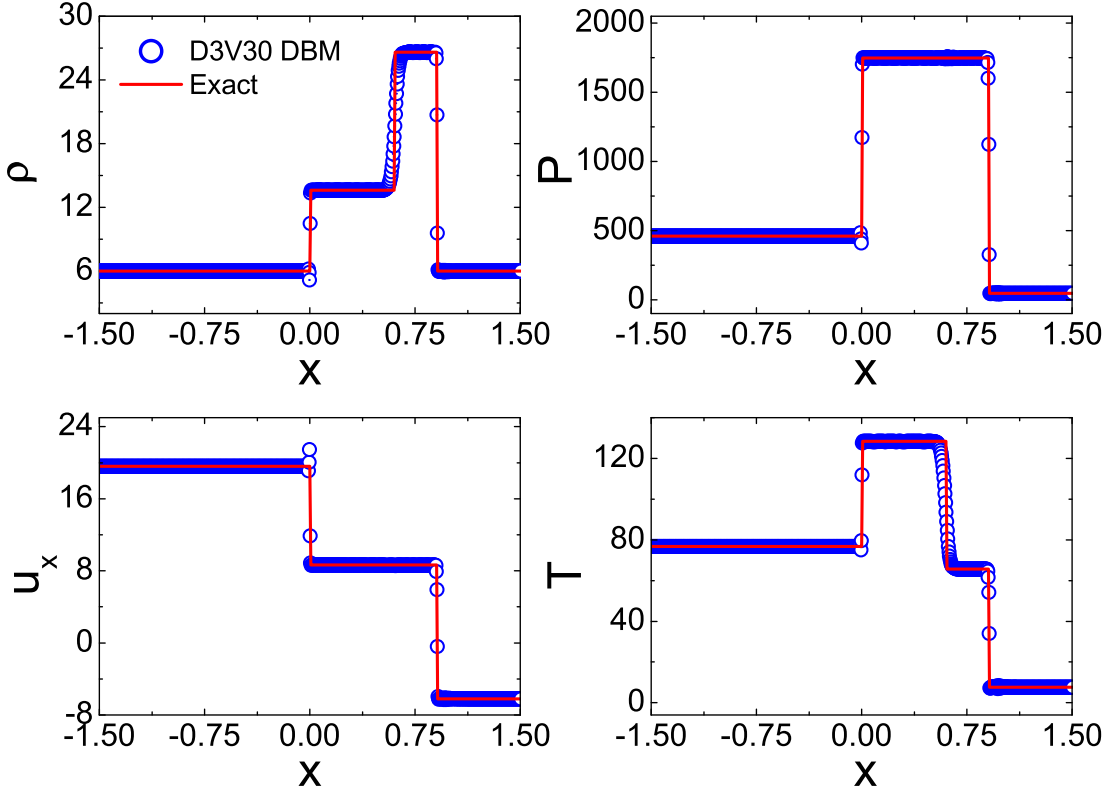,bbllx=6pt,bblly=3pt,bburx=575pt,bbury=412pt,
width=0.9\textwidth,clip=}}}
\caption{Comparisons between DB simulations and the exact solutions for the
problem of shock collision.}
\end{figure}

With respect to the former two tests, we consider a more challenging one,
i.e., the collision between two strong shocks, with the following conditions
\begin{equation*}
\left\{
\begin{array}{c}
(\rho ,T,u_{x},u_{y}, u_z)|_{L}=(5.99924,76.8254,19.5975,0.0, 0.0), \\
(\rho ,T,u_{x},u_{y}, u_z)|_{R}=(5.99242,7.69222,-6.19633,0.0,0.0).%
\end{array}%
\right.
\end{equation*}%
Exact solution of this problem includes a left and a right shocks, both
spreading to the right side, a contact discontinuity also moving rightwards.
Among which, the left shock propagating rightwards very slowly, which gives
rise to an additional challenge to the numerical algorithm. Figure 4 exhibits
simulation results from the D3V30 model, where $t=0.07$, $\gamma =1.5$.
Model parameters used here are $\Delta x=\Delta y=4\times 10^{-3}$, $\Delta
t=10 \tau=10^{-4}$, $c=15$, and $\eta _{0}=30$. Clearly, the simulations are
also in accordance with the theoretical solutions. Successful simulation of
this rigorous problem demonstrates that the model constructed by our
approach is robust and accurate enough to investigate compressible flows
with strong shock wave interactions. Of course, the unphysical oscillations
appear in profiles of $\rho$, $P$, and $u_x$ at around $x=0$ deserve further
consideration.

\subsubsection{High-Mach-number shock tube problem}

\begin{figure}[tbp]
{%
\centerline{\epsfig{file=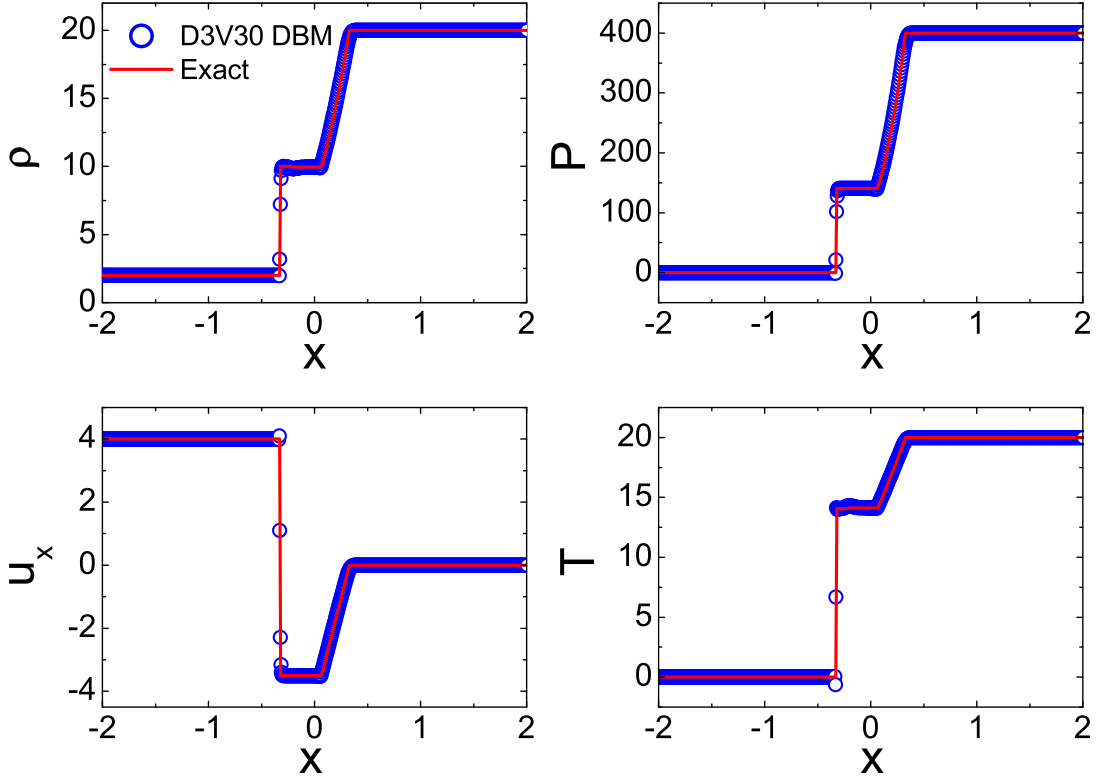,bbllx=6pt,bblly=2pt,bburx=581pt,bbury=407pt,
width=0.9\textwidth,clip=}}}
\caption{Comparisons between DB simulations and the exact solutions for the
high-Mach-number shock tube problem.}
\end{figure}

To further examine the adaptability of the model for high-Mach-number
problem, we propose a shock tube with the highest Mach number $Ma = 103.3$.
The initial conditions are
\begin{equation*}
\left\{
\begin{array}{c}
(\rho ,T,u_{x},u_{y},u_{z})|_{L}=(2.0,0.001,4.0,0.0,0.0), \\
(\rho ,T,u_{x},u_{y},u_{z})|_{R}=(20.0,20.0,0.0,0.0,0.0),%
\end{array}%
\right.
\end{equation*}
Figure 5 gives simulation results from D3V30 model and the exact solutions
at $t=0.06$. Parameters are selected as $\Delta x=\Delta y=\Delta z=5 \times
10^{-3}$, $\Delta t=2\tau=10^{-4}$, $c=6$, and $\eta_{0}=100$. Obviously,
the two sets of results coincide with each other. Moreover, there not exist
spurious oscillations near the sharp discontinuities. Successful simulation
of this test manifests the robustness, accuracy and adaptation for
compressible flows with high-Mach-number ($10^2$), high temperature ratio ($%
2\times10^4$) and high pressure ratio ($2\times10^5$).

\subsection{Two-dimensional Riemann problems}

\begin{table}[tbp]
\begin{tabular}{c||cc}
\hline\hline
configuration & $\left(
\begin{array}{c|c}
\begin{array}{cc}
P_{2} & \rho _{2} \\
u_{x2} & u_{y2}%
\end{array}
&
\begin{array}{cc}
P_{1} & \rho _{1} \\
u_{x1} & u_{y1}%
\end{array}
\\ \hline
\begin{array}{cc}
P_{3} & \rho _{3} \\
u_{x3} & u_{y3}%
\end{array}
&
\begin{array}{cc}
P_{4} & \rho _{4} \\
u_{x4} & u_{y4}%
\end{array}%
\end{array}%
\right) $ &  \\ \hline
$%
\begin{array}[b]{c}
\text{configurations} \\
\text{(a) and (b)}%
\end{array}%
$ & $\left(
\begin{array}{c|c}
\begin{array}{cc}
1 & 1 \\
0.7276 & 0%
\end{array}
&
\begin{array}{cc}
0.4 & 0.5313 \\
0 & 0%
\end{array}
\\ \hline
\begin{array}{cc}
1 & 0.8 \\
0 & 0%
\end{array}
&
\begin{array}{cc}
1 & 1 \\
0 & 0.7276%
\end{array}%
\end{array}%
\right) \left(
\begin{array}{c|c}
\begin{array}{cc}
0.35 & 0.5065 \\
0.8939 & 0%
\end{array}
&
\begin{array}{cc}
1.1 & 1.1 \\
0 & 0%
\end{array}
\\ \hline
\begin{array}{cc}
1.1 & 1.1 \\
0.8939 & 0.8939%
\end{array}
&
\begin{array}{cc}
0.35 & 0.5065 \\
0 & 0.8939%
\end{array}%
\end{array}%
\right) $ &  \\ \hline
$%
\begin{array}[b]{c}
\text{configurations} \\
\text{(c) and (d)}%
\end{array}%
$ & $\left(
\begin{array}{c|c}
\begin{array}{cc}
1 & 1 \\
-0.6259 & 0.1%
\end{array}
&
\begin{array}{cc}
0.4 & 0.5197 \\
0.1 & 0.1%
\end{array}
\\ \hline
\begin{array}{cc}
1 & 0.8 \\
0.1 & 0.1%
\end{array}
&
\begin{array}{cc}
1 & 1 \\
0.1 & -0.6259%
\end{array}%
\end{array}%
\right) $ \ \ $\left(
\begin{array}{c|c}
\begin{array}{cc}
0.4 & 0.5313 \\
0.8276 & 0%
\end{array}
&
\begin{array}{cc}
1 & 1 \\
0.1 & 0%
\end{array}
\\ \hline
\begin{array}{cc}
0.4 & 0.8 \\
0.1 & 0%
\end{array}
&
\begin{array}{cc}
0.4 & 0.5313 \\
0.1 & 0.7276%
\end{array}%
\end{array}%
\right) $ &  \\ \hline
$%
\begin{array}[b]{c}
\text{configurations} \\
\text{(e) and (f)}%
\end{array}%
$ & $\left(
\begin{array}{c|c}
\begin{array}{cc}
1 & 2 \\
0 & -0.3%
\end{array}
&
\begin{array}{cc}
1 & 1 \\
0.3 & 0%
\end{array}
\\ \hline
\begin{array}{cc}
0.4 & 1.0625 \\
0 & 0.2145%
\end{array}
&
\begin{array}{cc}
0.4 & 0.5197 \\
0 & -0.4259%
\end{array}%
\end{array}%
\right) $ \ \ $\ \left(
\begin{array}{c|c}
\begin{array}{cc}
1 & 2 \\
0.5 & 0.5%
\end{array}
&
\begin{array}{cc}
1 & 1 \\
0.5 & -0.5%
\end{array}
\\ \hline
\begin{array}{cc}
1 & 1 \\
-0.5 & 0.5%
\end{array}
&
\begin{array}{cc}
1 & 3 \\
-0.50 & -0.5%
\end{array}%
\end{array}%
\right) $ &  \\ \hline\hline
\end{tabular}%
\caption{Initial conditions for the 2D Riemann problems.}
\end{table}

In this subsection, the D3V30 DB model is adopted to solve the 2D Riemann
problems which own very rich and complicated wave patterns.
In detail, according to Refs. \cite{2D1,2D2,2D3}, $19$
genuinely different admissible configurations for polytropic gas, separated
by the shock wave, the rarefaction wave, and the contact wave, exist.
In the present study, we deal with the $6$ of which configurations with
the initial conditions illustrated in table II, where ($P_i, \rho_i, u_{xi}, u_{yi}$) denote
quantities in the $i$th quadrant.
The simulations are carried out on $400\times400\times2$ grid nodes.
Parameters are consistent with what we used in Fig. 2. In the $x$
and $y$ directions, the outflow boundary conditions are applied, in the $z$ direction, the PBCs are applied.
Figure 6 displays the simulated density patterns, where $80$
contours in panels (a) and (b), $40$ contours in panels (c)-(f).
As excepted, the computational results further verify the complexity of Riemann problems
for 2D gas dynamics. Generally, the contours in each case are smooth and continuous,
with high resolution and without overshoots or spurious numerical oscillations.
More importantly, the present kinetic model recovers the consistent
and detailed information on the variety of wave configurations,
compared to those by the traditional numerical methods \cite
{2D1,2D2,2D3,2D-A,2D-ABDF}, LB method \cite{Shu-IJNMF2010}, and gas-kinetic flux solver scheme \cite{Shu-JCP2015}.
Specifically, for configuration (a) at $t=0.1$, where initially possesses two shocks and two contact discontinuities,
the two slip lines $J_{32}$ and $J_{34}$ encountered with the sonic circles of the constant state in the third quadrant,
the pair of triple-shock-wave structure, together with the ripples in the first quadrant are
captured in accordance with scenarios in Ref. \cite{2D-A,2D-ABDF}.
Here $J_{ij}$ represents contact wave between the $i$th and $j$th quadrants.
Similarly, in configuration (b), a pair of three-shock configurations created by interactions between
the shocks $S_{21}$ and $S_{32}$, and an oval subsonic area are observed at $t=0.1$.
In configuration (c), slip lines $J_{32}$ and $J_{34}$ in the third quadrant are homologous with that in case (a).
Moreover, the unique semi-circular wavefront is reproduced with high resolution.
In configuration (d), the two shocks $S_{21}$, $S_{41}$, and two slip lines $J_{32}$ and $J_{34}$,
an oval subsonic region, and ripples in the first quadrant, are clearly presented.
In configuration (e), we observe that the whole domain is roughly divided into two equal portions by the
slip lines $J_{21}$ and $J_{34}$ with a cortex inside the subsonic region.
In the last configuration, the model reproduces a symmetric clockwise vortex which is composed of four
slip lines .

\begin{figure}[tbp]
\center{\epsfig{file=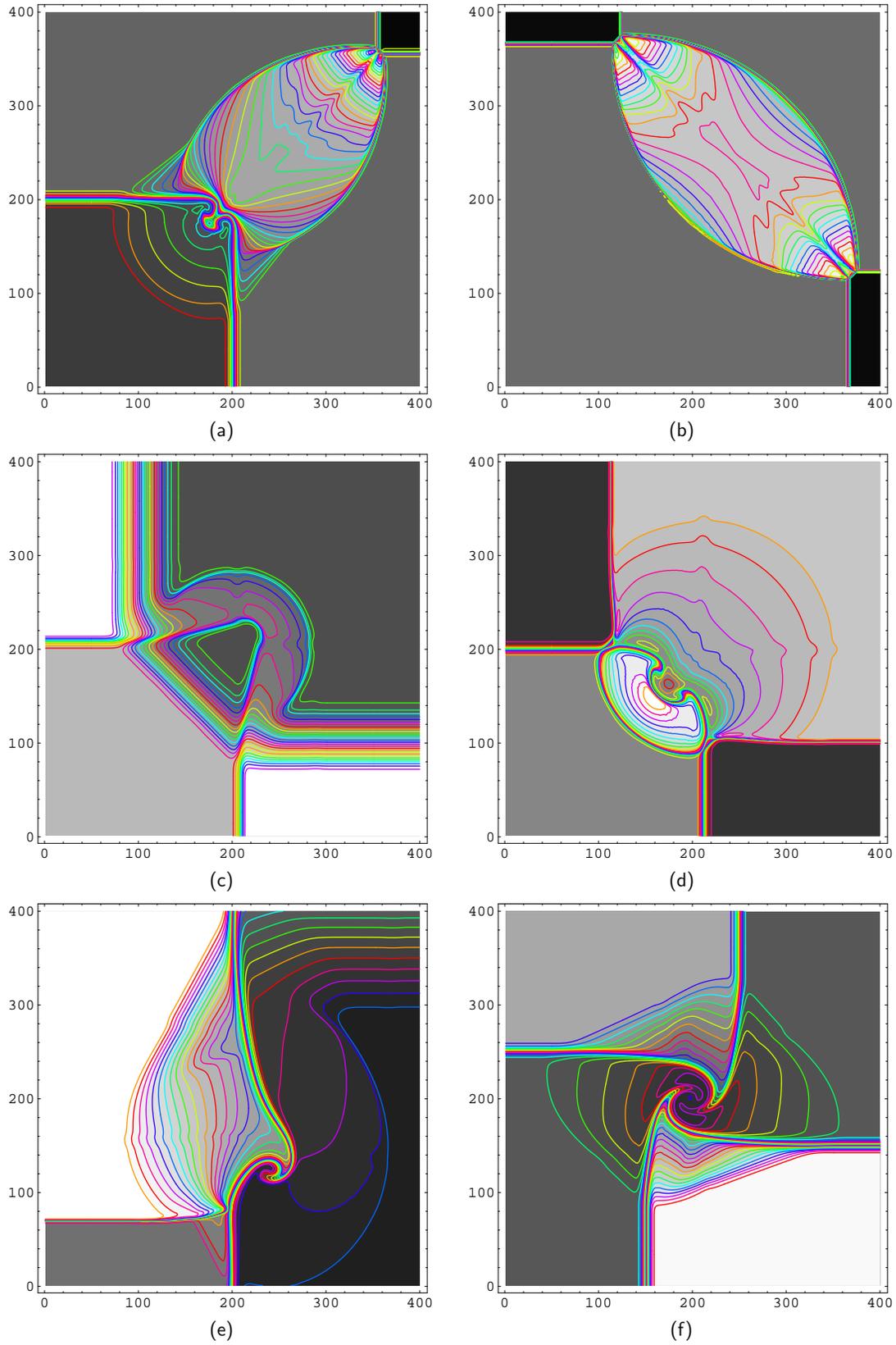,bbllx=106pt,bblly=18pt,bburx=569pt,bbury=708pt,
width=0.9\textwidth,clip=}}
\caption{ Density patterns calculated via the D3V30 model, where $80$
contours in panels (a) and (b), $40$ contours in panels (c)-(f). }
\end{figure}

\subsection{3D Riemann problems: Expanding shock in an open space}

\begin{figure}[tbp]
\center{\epsfig{file=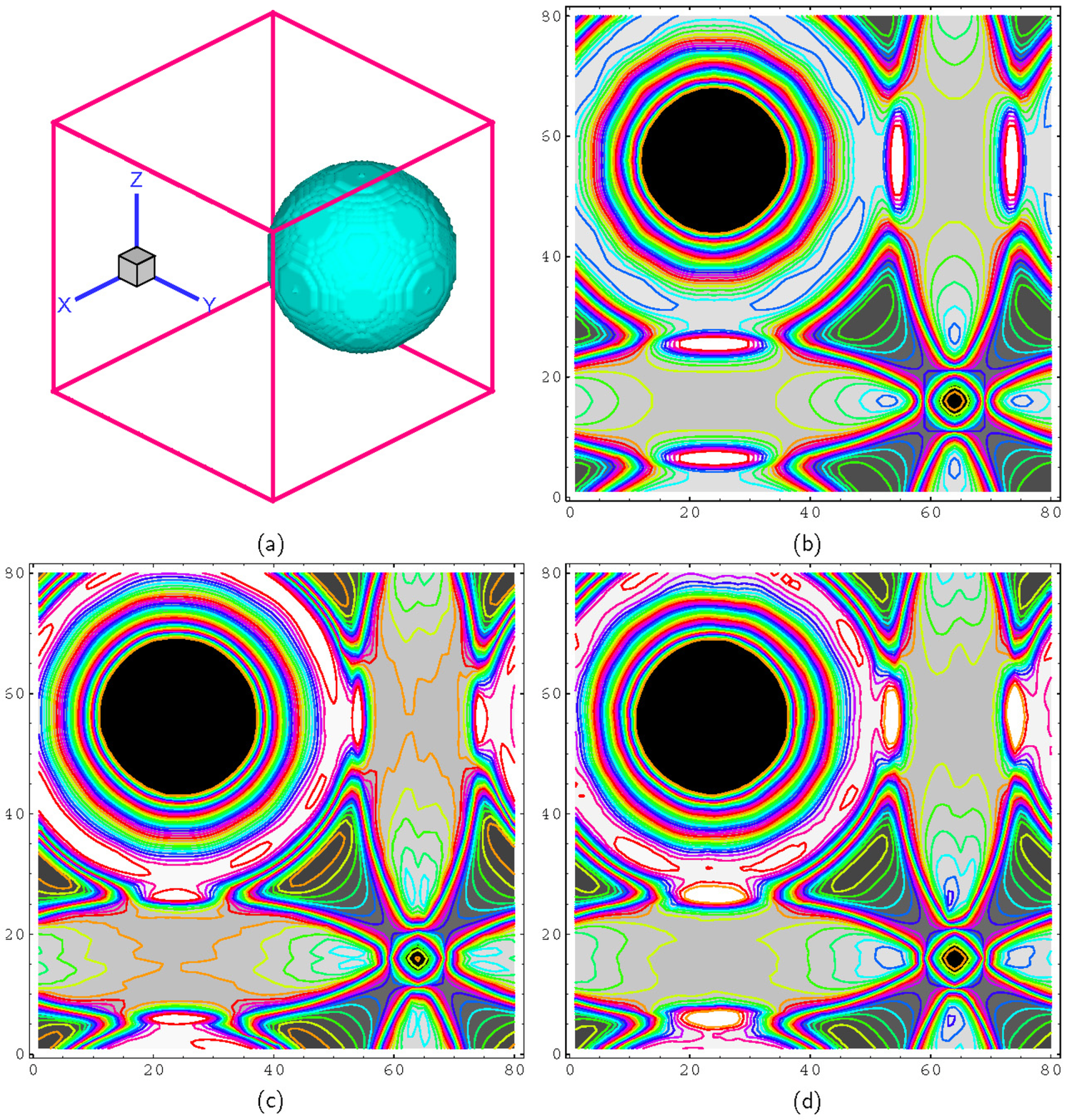,bbllx=106pt,bblly=206pt,bburx=575pt,bbury=696pt,width=0.94\textwidth,clip=}}
\caption{ Initial configuration of the 3D expanding shock in an open space (a). 40 density
contours at $i_{z}=N_{z}/4$ calculated from the D3V15 model (b), the D3V14 model (c), and the D3V30 model (d).}
\end{figure}

Here we consider a 3D test: expanding shock in an open space \cite{Web,Li-2009-PLA-3D,Li-2013-PA-3D}, i.e., a spherical shock wave expands in an open space.
Initially, the whole velocity is zero, the pressure and density are given
as $(\rho, P)=(5.0,5.0)$ when $\sqrt{(i_{x}-0.3N_{x})^2+(i_{y}-0.7N_{y})^2+(i_{z}-0.5N_{z})^2}\leq0.3N_{x}$;
otherwise $(\rho,P)=(1.0,1.0)$ with periodic boundary conditions at all
surfaces. A uniform mesh $N_{x} \times N_{y} \times N_{z} = 80 \times 80
\times 80$ is employed. Figure 7(a) shows the initial configuration of the
simulation. Parameters are unchanged with Fig. 6.
It should be noted that, the adaptation of the IMEX Runge-Kutta scheme makes the times step $\Delta t$ much larger than the relaxation time $\tau$. This is of great importance for 3D simulations where large amount of computation requires. Panels (b)-(d) in Fig. 7 exhibit the density contours at $i_{z}=N_{z}/4$, where (a), (b) and (c) are
obtained from the D3V15 model with additional physical viscosity \cite{3D-CTP},  the D3V14 model, and the D3V30 model, respectively.
Excellent agreement can be found among results from the three models.
The complex features such as the shocks and the interactions among the shocks
are properly captured. Nevertheless, it is necessary to point out that, the
model in Ref. \cite{3D-CTP} introduces some artificial viscosity to ensure
the numerical stability, and too much viscosity will broaden the width of the
shock wave, destroy the shock wave structure and pollute the whole velocity
fields. Therefore, how to introduce the viscosity as little as possible is
an important issue that remains consideration, while in the present models, this problem
does not exist.

\section{Conclusions and Remarks}

In this paper, a series of 3D DB models for compressible fluids in and out equilibrium are proposed.
The key formulating technique is the construction of DEDF through
inversely solving the kinetic moment relations that it satisfies.
Compared to the polynomial approach for formulating DEDF where DEDF
is expanded in terms of macroscopic variables with a global or distribute weighting coefficients, the present way is (i) conceptually simple, concise and straightforward, (ii) physically flexible and general, and (iii) numerically efficient.
DB models at the Euler and Navier-Stokes levels constructed by this approach have been verified by applications to some well-known benchmarks, ranging from 1D to 3D. The simulation results are in excellent agreement with Riemann solutions or results from computational fluid dynamics. The construction of higher-order DB models at the Burnett and super-Burnette levels via this approach is straightforward, and left for future study. The necessity of such a kinetic model is that, with increasing the complexity of flows, the dynamical characterization of non-equilibrium state and the understanding of the constitutive relations need the higher-order kinetic moments and their evolutions.

\section*{Acknowledgments}

The project is supported by the National Natural Science Foundation of China
(11475028, 11602162 and 11772064),
Science Challenge Project(under Grant No. JCKY2016212A501),
Natural Science Foundation of Hebei Province
(A2017409014 and A201500111), Natural Science Foundation of Hebei
Educational Commission (ZD2017001), and FJKLMAA, Fujian Normal
University.

\end{document}